\documentclass[11pt]{article}
\usepackage{amssymb,bm,a4}

\newcommand{\eeq}{\end{equation}}
\newcommand{\beq}{\begin{equation}}
\newcommand{\ba}{\begin{array}}
\newcommand{\ea}{\end{array}}
\newcommand{\bea}{\begin{eqnarray}}
\newcommand{\eea}{\end{eqnarray}}
\newcommand{\vev}[1]{\langle #1\rangle}

\newcommand{\preprintno}[1]{\vspace{-2cm}{\normalsize\begin{flushright}#1\end{flushright}}\vspace{1cm}}

\begin{document}

%\preprint{MCTP-02-30}

\title{\preprintno{\bf UT-STPD-2-03} 
Varying alpha, thresholds and extra dimensions}

\author{Thomas Dent\thanks{email:\,tdent@auth.gr}\\
	{\em Physics Division, School of Technology}\\
	{\em Aristotle University of Thessaloniki, Thessaloniki 54124, 
	Greece}\\
%\email[]{tdent@umich.edu}
%\thanks{email: tdent@umich.edu}
%\affiliation{Michigan Center for Theoretical Physics,\\
%University of Michigan, Ann Arbor, MI 48109}
}


\date{Spring 2003}

\maketitle

\begin{abstract}
\noindent We consider variations of coupling strengths and mass ratios
in and beyond the Standard Model, in the light of various mechanisms
of mass generation. In four-dimensional unified models, heavy quark 
and superparticle thresholds and the electron mass can completely alter
the (testable) relation between $\Delta \ln \alpha$ and $\Delta \ln \mu$, 
where $\mu\equiv m_p/m_e$. In extra-dimensional models where a 
compactification scale below the fundamental scale is varying, definite 
predictions may result even without unification; we examine some models 
with Scherk-Schwarz supersymmetry-breaking.
\end{abstract}

%\pacs{}
%\keywords{ }

%\maketitle

\section{Introduction}\noindent
If the recently measured cosmological variation in $\alpha$~\cite{Webb03}
is to be explained within a unified model, {\it i.\,e.}\/\ one in which all 
couplings are constrained by a relation at a certain energy scale, 
definite predictions should ensue for the variations of other quantities 
accessible to astrophysical measurement, in particular $\mu\equiv m_p/m_e$
\cite{Ivanchik} and $g_p$ (the gyromagnetic ratio of the 
proton)~\cite{CowieS,Murphy01}. The ``theory of everything'' which would be 
required to make such predictions does not at present exist, although there 
are candidates for large parts of it. So far, results have been somewhat 
discouraging for unification, even considering the amount of theoretical 
uncertainty attached~\cite{Calmet1,Langacker,us,Dine,PC_S_T}.

To summarize the situation very roughly, the mechanism(s) of mass generation 
in $d=4$ field theory, including dimensional transmutation in QCD as well 
as various ways of generating a Higgs v.\,e.\,v.\ and fermion mass 
hierarchies, may be very sensitive to changing couplings at high scale, 
since they can involve slow logarithmic running into the strong coupling 
regime. But the running of the U$(1)$ and SU$(2)$ gauge couplings is weak 
and, crucially, $\alpha$ does not depend exponentially on a gauge coupling 
at high scale, but only linearly (to first approximation).

Thus the fractional variations in the mass ratios which determine $\mu$ 
(and may affect $g_p$) obtain contributions many times the size of the
fractional variation of $\alpha$. However experimental bounds on variations 
in $\mu$ and $g_p$ are of a similar order to the signal in $\alpha$, 
namely a few times $10^{-5}$. Hence the number of theories that could 
possibly be consistent with experiment appears drastically reduced and may 
indicate a fine-tuning problem. Such a conclusion turns out to hold also 
if the parameters of the Higgs and flavour sectors are held fixed (relative
to the fundamental scale), 
%and the change at high scale is referred to a kind of threshold effect, 
%equivalent to RG evolution of the unified theory, 
since the variation of the SU$(3)$ coupling at high scale 
%arising through the beta-function of the unified theory 
is almost always at least as large as that of the U$(1)$ and SU$(2)$ 
couplings (whether the variation arises from a varying unified coupling
constant, or a varying GUT scale relative to the fundamental scale, or
some combination) and the exponential amplification due to strong running 
of QCD is always present~\cite{Calmet1,Langacker,us,Dine,PC_S_T}.

Our aim is not to cure this apparent problem, since it is obviously 
desirable that theories be ruled out by experiment. Instead we will 
take more fully into account effects associated with thresholds in RG 
running and mass generation, and show that they may significantly 
alter theoretical expectations, and also point out that predictions can 
be made in the case of a varying compactification radius even in models
without unification. 

Important thresholds arise already in the SM, for example from the 
$W^\pm$ and heavy quarks; theories capable of addressing the hierarchy
problem are expected to possess additional thresholds, and imply particular 
relations for variations in the mass spectrum (which, of course, directly 
influences $\mu$). To make predictions, one must identify a single varying 
degree of freedom in the model as the source of the observed effect: given 
the complexity of realistic unified theories there may be more than one 
candidate for the source of the variation. Nevertheless certain 
possibilities are obviously simpler and better-motivated than others.

For a varying GUT gauge coupling, given the SUSY-GUT value 
$M_G\simeq 2\times 10^16\,$GeV and without considering variation of mass 
ratios in the electroweak and flavour sectors~\cite{Calmet1,Langacker,us}, 
earlier calculations gave 
$\bar{R} \equiv \Delta \ln \mu/\Delta \ln \alpha = 36$ with an 
uncertainty of a few due to higher-order effects. Our estimates, including 
varying mass ratios~\footnote{In the case when electroweak symmetry-breaking 
occurs in a hidden sector with a gauge coupling varying with the unified 
coupling}, and with an improved treatment of the proton 
mass~\footnote{See also~\cite{FlambaumS,Wetterich02}} give 
$\bar{R} = -13 \pm 7$ in a theory with softly-broken supersymmetry at
the weak scale and $\bar{R} = 4 \pm 5$ without. 

\subsection{New variations and twists from extra dimensions}
Various types of model with extra dimensions
\cite{Antoniadis,DDG,Quiros,Nomurawk,Kawamura,Nomuragut,Nomuradesert} 
may have interestingly different predictions since first, gauge and matter 
fields may propagate in different submanifolds in extra dimensions; second, 
the RG scale at which the variation of couplings originates is not tied to 
any unification scale; and lastly, the mechanisms of breaking 
SU$(2)\times$U$(1)$, supersymmetry or GUT groups in some theories, have a 
distinctive dependence on the radii of compact extra dimensions. 

On the first point, traditional Kaluza-Klein-like theories 
have a universal dependence on the size of the extra dimension, since all 
fields live on the same manifold; however, ``brane-world'' 
theories restrict some fields to a submanifold. The submanifold may 
indeed only be $R^3\times S^1$, in which case couplings involving localized
fields are not affected when the size of transverse dimensions varies. Of 
course, if all fields live on the same brane, there is no new behaviour 
of varying couplings (except for gravity) resulting from the presence of
extra dimensions.

On the second point, the couplings need no longer be determined (by, say, 
a dilaton v.\,e.\,v.) at GUT or string energy scales and run through 14 or 
more orders of magnitude to reach experimentally-accessible energy scales. 
Both the fundamental cutoff scale, and compactification scales, can be much 
lower, thus RG running may be quantitatively different: the scale at which 
the variation in couplings originates is different. This possibility 
includes the case of extra-dimensional GUTs with Kaluza-Klein thresholds 
\cite{DDG,PC_S_T}. In fact, even non-GUT theories may have 
well-defined predictions, in the case where the variation is due to a
cosmologically varying radius. We only need to know the physics at and 
below the energy scale where the variation is transmitted to the SM 
fields: in general couplings at this scale will not be unified. (Compare 
\cite{ChackoGP}, where the unification relations are irrelevant for the 
effect discussed).

On the third point, Scherk-Schwarz (S-S) symmetry breaking 
\cite{S-S,Antoniadis,Quiros,Nomurawk} generates a mass scale inversely 
proportional to the radius of compactification, while perturbative
couplings in the effective $D=4$ theory vary as a power of the radius 
(depending on which fields propagate in extra 
dimensions~\footnote{Classically, the 4d coupling constant of a gauge field 
$\alpha_i^{(4)}$ propagating in extra dimensions varies simply as $R_i^{-1}$, 
where $R_i$ is the radius associated with a one-cycle of a compactification 
manifold, which in the simplest case has the metric of a torus.}).
Thus for a varying radius, mass ratios and coupling strengths (defined
at the scale of the inverse radius) will both vary as power laws with small 
exponents. This contrasts with the exponential dependence of mass scales on 
the high-energy gauge coupling in the case of mass generation by dimensional 
transmutation in $D=4$.~\footnote{One cannot alter the exponential dependence 
of $\Lambda_c$, the confinement scale of strong interactions, on the 
asymptotic SU$(3)$ coupling constant without radically altering 
phenomenology.}

\subsection{Other issues}\label{sec:other}
There are many aspects of varying couplings that we will not discuss in 
detail. One outstanding question is what form the time- (or space-) 
dependence takes and what dynamics generate it. Following on from this 
question, if the variation originates from a light scalar, long-range 
forces and apparent violations of the weak equivalence principle will likely 
ensue~\cite{DvaliZ,Wetterich02,DamourPV}: see also Section \ref{sec:longr}. 

If the spacetime dependence could be predicted, then bounds from the Oklo 
phenomenon~\cite{DamourD,Fujii} and from the early Universe (in particular 
nucleosynthesis and CMB)~\cite{us,Martins,FlambaumS} might well be the 
most stringent.\footnote{These bounds, which depend on nuclear or 
gravitational physics, do not directly limit $\alpha$, rather they probe 
some combination of $\alpha$ and other dimensionless couplings: thus one can 
imagine degeneracies, enhancements~\cite{Olive_et_al,us} or even 
cancellations depending on how other couplings may vary.} We will take 
the opposite approach and concentrate on the range of redshifts at which 
a nonzero variation in alpha is claimed: direct constraints arise 
from astrophysical observations at the same epoch. 
See~\cite{Langacker,Calmet2,Dine,PC_S_T} for similar approaches.

The case of spatial variation~\cite{BarrowOT} is less straightforward. 
Binning points by redshift is then not useful, and the dependence of 
alpha on the source (actually, on the absorbing medium) could be 
highly model-dependent. It could be that couplings measured on Earth 
are different from those in the gas clouds which are under observation
simply by virtue of the different astrophysical environment --
possibly due to the different gravitational potentials~\cite{Rafelski,Mbelek}.

Whatever the dynamics producing the variation, one can investigate 
correlations between (variations in) the quantities that determine the 
spectral lines, $\alpha$, $g_p$ and $\mu$. That these quantities are not 
measured simultaneously for the same sources or absorbers may be of 
concern. In models with variation over cosmological time, we have to assume 
that the variation is smooth and relatively slow. In models with spatial 
variation, we might expect a correlation between the variation of $\alpha$ 
in one source or absorber, relative to on Earth, and variations of $g_p$ 
and $\mu$ in another; however in the case of strongly environment-dependent
variation we would need much more data on the latter
quantities, which are currently bounded by a small handful of systems. If the 
spatial variation is long-range, rather than environment-dependent, one can 
say little without more detailed work on the spatial distribution of 
the sources. Spatial variation might be diagnosed by measurements at low 
redshift, or by analysis of spatial correlations (requiring a much larger
data set than at present).

The problem of fine-tuning the mass of any scalar field responsible for the 
variation against radiative corrections is an acute one, as noted in 
\cite{Choi} (in the context of quintessence) and~\cite{BanksDD}: we have no 
further insights into this. The possibility noted in~\cite{ChackoGP} of a 
rather heavier scalar $m\sim 10^{-3}\,$eV in a new weakly-coupled sector 
alleviates such a problem, but the gravitational energy of the scalar
must be dynamically cancelled if the model is to be cosmologically viable.
%(There may also be problems with nucleosynthesis since new light degrees of 
%freedom are introduced, with an energy density comparable to neutrinos;
%recent data disfavour such scenarios~\cite{CrottyLP}).

Finally, the possibility of degeneracy in the interpretation of
measurements in terms of atomic physics has been raised~\cite{Calmettalk}: 
the changes in spectral lines interpreted as varying alpha could in 
principle be due to varying $\mu$ or $g_p$ instead. A full evaluation for 
the many-multiplet technique could be complicated, as one needs to know 
the dependence of many spectral lines on all three parameters; recall that 
numerical calculations were required to find the expected variations in 
frequency due to alpha alone.
% except if there were compelling reasons to rule out one or the other possibility. 
We will, for the moment, take the data on $\alpha$, $\mu$ and $g_p$ at face 
value.

\subsection{Note added}
After the first version of this paper appeared, Murphy {\em et al.} 
released a preprint \cite{Webb03} with a revised value of $\Delta 
\alpha/\alpha$, with the change being due to the identification of an 
additional source of random scatter in some of the systems studied, and a 
corresponding increase in the statistical errors assigned. This new value
is used in the discussion in Section \ref{sec:data}.

\section{Mass generation and thresholds}
The importance of the Higgs v.\,e.\,v.\ (or the ratio $v_H/M_P$) was realised 
in~\cite{C+O}, which found that its dependence on the SUSY-breaking masses 
and the top Yukawa was dominant in determining the outcome of 
nucleosynthesis. We will find that the variation of $v_H$ (and superpartner 
masses) can also be decisive for the spectral data, since they depend 
crucially on mass ratios, particularly the ratios of quark and lepton masses 
to the scale $\Lambda_c$ of QCD. After a general discussion of how 
thresholds enter the low-energy observables, we give examples of how the 
inclusion of thresholds, formally a higher-loop effect, can completely alter
predictions from unified theories.

We fix notation by quoting the (solution of the) one-loop RG equation 
for gauge couplings:
\beq \label{eq:1}
\alpha_i^{-1}(\mu^-) = \alpha_i^{-1}(\mu^+) - \frac{b_i}{2\pi} \ln\left(
\frac{\mu^-}{\mu^+}\right),
\eeq
thus $\beta_i$ is negative for asymptotically free groups. Now given a 
charged field whose decoupling mass is $m^{\rm th}(m^{\rm th})$ (in a 
convenient mass-independent renormalisation scheme), we have
\beq \label{eq:2}
\alpha_i^{-1}(\mu^-) = \alpha_i^{-1}(\mu^+) - \frac{b^-_i}{2\pi} \ln\left(
\frac{\mu_-}{\mu^+}\right) - \frac{b_i^{\rm th}}{2\pi}
\ln\left(\frac{m^{\rm th}}{\mu^+}\right)
\eeq
where $b_i^{\rm th}\equiv b_i^+ -b_i^-$, the beta-function coefficient being 
$b_i^+$ above the threshold and $b_i^-$ below, with tree-level matching at
$m^{\rm th}$. In the case of multiple thresholds one sums the corrections
to $\alpha_i^{-1}$. For the QCD invariant scale $\Lambda_c\equiv M 
e^{-2\pi/9\alpha_3(M)}$ where $m_s<M<m_c$ we find for non-supersymmetric 
theories
\beq \label{eq:3}
\frac{\Lambda_c}{\mu^+} = e^{-2\pi/9\alpha_3(\mu^+)} 
\left(\frac{m_c m_b m_t}{\mu^{+3}}\right)^{2/27}
\eeq
where $\mu^+ >m_t$, and for superpartners (squarks of geometric average 
mass $m_{\tilde{q}}$ and gluinos of mass $m_{\tilde{g}}$)
\beq \label{eq:4}
\frac{\Lambda_c}{\mu^+} = e^{-2\pi/9\alpha_3(\mu^+)} 
\left(\frac{m_c m_b m_t}{\mu^{+3}}\right)^{2/27} 
\left(\frac{m_{\tilde{q}} m_{\tilde{g}}}{\mu^{+2}}\right)^{2/9} 
\eeq
where $\mu^+ >m_{\tilde{q}}m_{\tilde{g}}$ and the powers $2/27$ and $2/9$ 
come from ratios of beta-function coefficients, specifically 
$-b_3^{\rm th}/b_3^{(<m_c)}=b_3^{\rm th}/9$. Thus the variation in 
$\Lambda_c$ is
\beq \label{eq:5}
\frac{\mu^+}{\Lambda_c}\Delta \frac{\Lambda_c}{\mu^+} = \frac{2\pi}
{9\alpha_3(\mu^+)} \frac{\Delta \alpha_3(\mu^+)}{\alpha_3(\mu^+)} +
\frac{2}{27}\sum_{q\,=\,c,b,t}\frac{\mu^+}{m_q}\Delta\frac{m_q}{\mu^+} 
 + \frac{2}{9} \left( \frac{\mu^+}{m_{\tilde{q}}} 
\Delta\frac{m_{\tilde{q}}}{\mu^+} + \frac{\mu^+}{m_{\tilde{g}}} 
\Delta\frac{m_{\tilde{g}}}{\mu^+} \right)  
\eeq
where terms in brackets are to be ignored for the non-supersymmetric case.
The heavy quark masses to be used are strictly the decoupling masses 
$m_q(m_q)$, but we showed in~\cite{us} that the difference between using 
this definition and the ``invariant quark mass'' $\hat{m}_q$ defined such
that it has no scale dependence, is negligible.

We see immediately that the threshold terms are of higher order in 
$\alpha_3$, thus formally they ought to be grouped with the power-law
correction to $\Lambda_c$ from two-loop running. However, when the Higgs 
v.\,e.\,v.\ is varying rapidly compared to $\alpha$, as is generic in 
theories unified at high scale~\cite{Langacker,us}, the threshold terms 
can become dominant, in contrast to the two-loop term, which is a
model-independent infrared effect. 

\subsection{Electroweak symmetry-breaking}
One case where thresholds are important is when electroweak 
symmetry-breaking (EWSB) is triggered by nonperturbative gauge theory 
effects and the gauge group giving rise to such effects is unified with 
the SM gauge couplings at high scale. This case encompasses gravity- and 
gauge-mediated SUSY-breaking and technicolor or composite theories, under 
the assumptions that all gauge groups be unified, and all mass scales 
lower than the fundamental scale $M_X$ (which may be the GUT scale, the 
string scale related to $\alpha'$ or the Planck scale of whatever underlying 
theory we are considering) are generated dynamically. The generic 
dependence of $v_H$ 
%on the unified coupling $\alpha_X$ 
is 
\beq \label{eq:6}
\frac{v_H}{M_X} = k \alpha_X^n e^{-2\pi m/b_h\alpha_X}
\eeq
where $k$ is a numerical constant and $n$ parameterises a possible 
power-law dependence, for example through the Higgs quartic coupling which
may (as in the MSSM) be related to a gauge coupling varying with $\alpha_X$,
or through the gravitational constant in the case of gravity-mediated 
SUSY-breaking. The value of $m$ is model-dependent, however the product 
$m/b_h\alpha_X$ can be estimated more reliably.
In the case of theories without fundamental Higgs, this equation simply 
parameterises the v.\,e.\,v.\ of whatever condensate breaks 
SU$(2)\times$U$(1)$. Thus,
\beq \label{eq:7}
\frac{M_X}{v_H} \Delta \frac{v_H}{M_X} \simeq \left(n + 
\frac{2\pi m}{b_h\alpha_X}\right) \frac{\Delta \alpha_X}{\alpha_X}.
\eeq
Now under the assumption that $k$ and $n$ are order 1, we neglect (the
logarithm of) their contribution to Eq.~(\ref{eq:6}) and find 
\[ \frac{2\pi m}{b_h\alpha_X} \simeq \ln[M_X/v_H \simeq 
(2\times 10^{16})/(2\times 10^2)] \simeq 32 
\]
where we identified $M_X$ with the SUSY-GUT value.~\footnote{In the 
non-supersymmetric case, single-step unification is of course inconsistent 
with data. With the nominal values $M_G\simeq 10^{15}\,$GeV and
$\alpha_G^{-1}\simeq 42$ we obtain $29$ rather than $32$ for this
equation. Non-SUSY unification is of course possible with intermediate
breaking scales, however such theories lose predictivity. In what follows 
the ``non-supersymmetric'' case will just mean the result when 
superpartner thresholds are ignored.}
Thus in Eq.~(\ref{eq:6})
the RHS becomes approximately $(n + 32) \Delta \alpha_X/\alpha_X$, which
is consistent with neglecting $n$ of order 1. If instead we take the 
heterotic string scale $M_X\sim 4\times 10^{17}\,$GeV we evaluate the
RHS as $(n + 35)$, thus the uncertainty introduced by a small integer 
value of $n$ can be neglected. For illustrative purposes we will take 
$\Delta \ln (v_H/M_X) = (34\pm 2) \Delta \ln \alpha_X$. 
From standard GUT relations, neglecting the effect of thresholds on the
running of SU$(2)$ and U$(1)_Y$ one finds~\cite{Calmet1,Langacker} 
$\Delta \ln \alpha \equiv \Delta \ln \alpha_{\rm em}(m_e) \simeq
0.5 \Delta \ln \alpha_X$, thus we may also write 
$\Delta \ln (v_H/M_X) = (68\pm 4) \Delta \ln \alpha$.\footnote{See 
section~\ref{sec:detailed} for a treatment including thresholds.}
Similarly, we may parameterise SUSY-breaking masses as 
\beq \label{eq:8}
\frac{\tilde{m}}{M_X} = k' \alpha_X^{n'} e^{-2\pi m/b_h\alpha_X} 
\eeq
where $m_{\tilde{q}}$, $m_{\tilde{g}}$ {\em etc.}\/\ vary as $\tilde{m}$, 
thus their variation can also be estimated as $\Delta \ln (\tilde{m}/M_X) 
= (68\pm 4)\Delta \ln \alpha$ given that superpartners are around the 
electroweak scale. 

It has been pointed out~\cite{RobertsRoss,C+O} that radiative EWSB may
depend sensitively on a combination of the top Yukawa coupling and
ratios of SUSY-breaking soft mass terms. Hence, any relation between 
$v_H/M_X$ and $\tilde{m}/M_X$ may be significantly more complicated. 
Variations in $y_t$ and soft term ratios are exceedingly model-dependent, 
to the extent that it is futile to look at detailed models. 
The best we can do is to parameterise our ignorance (somewhat as in 
\cite{Langacker}) by allowing the Higgs v.\,e.\,v.\ to have a different 
fractional variation $\Delta \ln (v_H/M_X)/\Delta \ln \alpha_X$ compared
to that of the average superpartner mass Eq.~(\ref{eq:8}). We put 
\[ 
\beta_v \equiv \frac{\Delta \ln (v_H/M_X)}{\Delta \ln \alpha_X},\ \ 
\beta_S \equiv \frac{\Delta \ln (\tilde{m}/M_X)}{\Delta \ln \alpha_X}
\]
where $\beta_v$ is defined for all models whether SUSY or non-SUSY and 
$\beta_S$ only in the supersymmetric case. Thus, from the above discussion,
if there is high-scale unification and EWSB is generated by dimensional
transmutation, we expect $\beta_v \simeq \ln(M_X/v_H)$, which also holds 
for the SUSY case in the absence of strong dependence on $y_t$ and soft 
mass ratios; if SUSY-breaking is triggered by nonperturbative effects in 
a hidden sector we have $\beta_S \simeq \ln(M_X/\tilde{m})$.

If the hidden sector group is not unified with the GUT group, in 
the sense of having the same coupling constant at $M_X$, these expectations
may be altered. In some heterotic string models one obtains gauge kinetic
functions $S+\varepsilon T$ for the visible sector and $S-\varepsilon T$ for the
hidden sector, where $S$ is the (four-dimensional) dilaton and $T$ a volume
modulus, and $1>\varepsilon >0$. If we identify $S$ as the varying d.\,o.\,f.\ 
the model behaves similarly to the case of a ``unified hidden sector'';
if we identify $T$ then the values of $\beta_S$ and $\beta_v$ will be 
large and negative! Hence, we keep $\beta_S$ and $\beta_v$ as free
parameters until the final stage of calculation.

\subsection{Initial estimate of $\mu$}
In the first approximation we neglect variations in Yukawa couplings
at high scale, and obtain
%proceed to estimate the variation in $\Lambda_c/M_X$:
\bea \label{eq:9}
\Delta \ln \frac{\Lambda_c}{M_X} &\equiv& \beta_\Lambda \Delta \ln 
\alpha_3(M_X) = \left(\frac{2\pi}{9\alpha_3(M_X)} + \frac{2}{9}\beta_v 
 + \frac{4}{9}\beta_S \right) \Delta \ln \alpha_3(M_X) \nonumber \\
% &=& \left( 17\pm (\mbox{2-loop}) + 
%\frac{2}{9}(34 \pm 2) 
%\left[ + \frac{4}{9}(\beta_S = 34 \pm 2) \right] \right)
%\frac{\Delta \alpha_X}{\alpha_X} \nonumber \\ 
 &=& (24 \pm{\rm few}) \frac{\Delta \alpha_X}{\alpha_X}\ \mbox{[non-SUSY],}\ \
 (39 \pm{\rm few}) \frac{\Delta \alpha_X}{\alpha_X}\ {\rm[SUSY]}
\eea
using $1/\alpha_X\simeq 24$, where we estimate ``few'' as about $3$ and 
set $\beta_S=0$ for a theory without superpartners. If the proton 
mass is well-approximated by a constant times $\Lambda_c$, we can 
estimate
\beq \label{eq:10}
\frac{\Delta \mu}{\mu} 
%= \left(\beta_\Lambda - \beta_v \right) \frac{\Delta \alpha_X}{\alpha_X}
= (-10 \pm {\rm few}) \frac{\Delta \alpha_X}{\alpha_X} \ \mbox{[non-SUSY],}\ \ (5 \pm {\rm few}) \frac{\Delta \alpha_X}{\alpha_X} \ {\rm[SUSY]}.
\eeq
The dependence of the electron Yukawa coupling may also come into the
estimate, possibly at the level of changing this value to, say, $-13[+2]$ 
if we imagine $y_e\propto \alpha_X^3$ due to some dynamics of flavour 
structure.
This expectation contrasts with the cases where the Higgs v.\,e.\,v.\ 
and superpartner masses are fixed relative to $M_X$, giving 
$\Delta\mu/\mu \simeq 17 \Delta\alpha_X/\alpha_X$, or where $v_H$ varies 
as in (\ref{eq:7}) but the resulting effects on and of QCD thresholds are 
neglected, giving $\Delta\mu/\mu \simeq -17 \Delta\alpha_X/\alpha_X$. 
Thus the mechanisms of mass generation and the presence of thresholds,
especially superpartner thresholds, can completely alter expectations 
for the relations between varying $\mu$ and $\alpha_X$. In the next 
section we look at the observables $\alpha$ and $\mu$ in more detail
and obtain preciser estimates taking mass generation and threshold effects
fully into account.

\subsection{More detailed treatment of $\alpha$ and $m_p$}\label{sec:detailed}
\subsubsection{$\alpha$ and thresholds}
At first sight, ``varying the unification scale'' and ``varying the 
electron mass'' while keeping other parameters ``constant'' should 
produce equivalent effects on $\alpha$: only the ratio of two masses is 
physical. However, the behaviour of intermediate thresholds is most 
important to determine the result. Hence, we include all known charged 
thresholds to determine what effect they may have on na{\" \i}ve GUT 
predictions.

``Varying $M_G$'' is interpreted as all low-energy threshold masses
(that originate from EWSB) varying by the same fraction with respect to 
$M_G$: thus $m_\mu/M_G$, $m_\tau/M_G$, $m_W/M_G$, {\em etc.} vary by the 
same fraction as $m_e/M_G$.~\footnote{Since the GUT scale is defined through the masses of superheavy fields that trigger the breaking to $G_{\rm SM}$, their threshold masses are constants in GUT units. However, see~\cite{Dine} and the discussion preceding Section 3.}
``Varying $m_e$'' is interpreted as holding all threshold masses fixed 
in units of $M_G$ except for $m_e/M_G$ itself. The beta-function 
coefficient per Dirac fermion of charge $Q$ is $4Q^2/3$, thus ``varying 
$m_e$'' leads to the relation
\[
\alpha_i^{-1} = \alpha_i^{-1}(\mu^+) 
%- \frac{b^-_i}{2\pi} \ln\left(\frac{\mu_-}{\mu^+}\right) 
 - \frac{(4/3)}{2\pi}\ln\left(\frac{m_e}{\mu^+}\right), \ \ 
 \frac{\Delta \alpha}{\alpha} = \frac{2\alpha}{6 \pi} 
 \Delta \ln\frac{m_e}{\mu^+} 
 \approx (7.7 \times 10^{-4}) \Delta \ln\frac{m_e}{\mu^+}
\]
where $m_e < \mu^+$ and we take $\alpha_i^{-1}(\mu^+)$ to be fixed.

Since we have $\alpha^{-1}=\alpha_1^{-1}+\alpha_2^{-1}$ at the weak
scale and ${\rm Tr}\, Q^2 = {\rm Tr}\, T_3^2 + Y^2 {\rm Tr}\, \bf{1}$ 
over an SU$\,(2)$ irrep, we easily see that the effect on $\alpha$ of a 
certain fractional variation in the masses of a SU$\,(2)$ multiplet with
mass below $M_W$ is no different from the same fractional variation in a 
multiplet lying above $M_W$.
% (using a geometric average mass for split multiplets). 
We have in general
\beq \label{eq:11}
\frac{\Delta \alpha}{\alpha} = \sum_{i=1,2}\frac{\alpha}{\alpha_i(\mu^+)}
 \frac{\Delta \alpha_i(\mu^+)}{\alpha_i(\mu^+)}
 + \alpha \sum_{\rm th} \frac{{Q^{\rm th}}^2 f^{\rm th}}{2\pi} 
%\frac{\mu^+}{m^{\rm th}} 
 \Delta \ln \frac{m^{\rm th}}{\mu^+}
\eeq
where $\alpha_1$ = ${g'}^2/4\pi$, the
second sum is over all charged fields, $f^{\rm th}$ is $2/3$ per chiral 
(or Majorana) fermion, $1/3$ per complex scalar and $11/3$ per vector boson.
All logarithms $\ln (m^{\rm th}/\mu^+)$ are numerically comparable if 
${\mu^+}$ is very large, however in the case of low-scale models the 
difference between $\ln (m_e/M_X)$ and $\ln (m_W/M_X)$ may be significant.
For the light quarks, the confinement scale $\Lambda_c$ provides a dynamical
cutoff. We have, setting aside the first term in
Eq.~(\ref{eq:11}),
\bea \label{eq:12}
\frac{\Delta \alpha}{\alpha}|_{\rm th}
 &=& \frac{1}{137\cdot 2\pi} \left( \frac{8}{3} \beta_\Lambda 
\Delta \ln \alpha_X
% \Delta \ln \frac{\Lambda_c}{M_X} 
+ \frac{4}{3} \left(\frac{4}{3} \Delta
 \ln \frac{m_cm_t}{M_X^2} + \frac{1}{3} \Delta \ln \frac{m_b}{M_X} 
 + 3\Delta \ln \frac{m_l}{M_X} \right) \right.\nonumber \\
 &-& \left. \frac{21}{3} \Delta \ln \frac{M_W}{M_X} \
  + 8 \Delta\ln \frac{\tilde{m}}{M_X} + \frac{1}{3} \Delta\ln 
 \frac{m_H}{M_X} \right) \\
 &=& \frac{1}{137\cdot 2\pi} \left( \frac{8}{3} \beta_\Lambda 
 + \beta_v + \frac{25}{3}\beta_S \right) \Delta \ln \alpha_X \label{eq:13}
\eea
where $m_l$ is an averaged lepton mass, we include the SM Higgs 
contribution as the longitudinal modes of the $W$'s and count the 
remaining modes of the second Higgs doublet in SUSY models as if 
superpartners. $\tilde{m}$ and $m_H$ stand for geometric averages over
superpartners and heavy Higgses respectively, where such fields exist.

Now in the case where $\alpha_X$ varies at a high scale, taking $\beta_v 
\simeq \beta_S = 34\pm 2$ and using the dependence of $\Lambda_c$ from 
Eq.~(\ref{eq:9}) both with and without SUSY thresholds we find
\[ \frac{\Delta \alpha}{\alpha}|_{\rm th} \simeq 
(0.11\pm 0.01) \frac{\Delta\alpha_X}{\alpha_X}\ \mbox{[non-SUSY]},\
(0.49 \pm 0.03) \frac{\Delta\alpha_X}{\alpha_X}\ \mbox{[SUSY]}
\]
which is a non-negligible correction to the direct contribution from 
varying $\alpha_{1,2}(M_X)$, $(\Delta\alpha/\alpha)_{\rm direct} = 
(8\alpha/3\alpha_X) \Delta\alpha_X/\alpha_X 
\simeq 0.47 \Delta\alpha_X/\alpha_X$. 
(With the the non-SUSY SU(5) estimate $\alpha_X\simeq 1/42$, this ratio 
is $0.82$.)

\subsubsection{$m_p$ contributions from light quarks and electromagnetism}
The proton mass receives several contributions: studies in chiral 
perturbation theory~\cite{BorasoyM} indicate a dominant contribution 
from nonperturbative effects in the chiral limit, which can only be 
proportional to $\Lambda_c$, and subleading terms depending on $u$, $d$, 
and $s$ quark masses. Heavy quarks are assumed to be well-described by 
their threshold effects on $\Lambda_c$. There is also an electromagnetic 
contribution~\cite{GasserL,MullerMeissner}. 

The effect of varying light quark masses can be found by applying the
Feynman-Hellmann theorem:
\[
\frac{m_q}{m_p} \Delta \frac{m_p}{m_q} \approx \frac{m_q}{m_p} 
\frac{\partial m_p}{\partial m_q} = m_q \vev{p|\bar{q} q|p}
\]
where the matrix element is essentially the so-called $\sigma$-term. 
Quark masses appear in this equation such that any multiplicative change 
of normalization of the $q$ operators cancels out: physical quantities do not
depend on a particular definition of quark masses. Hence, as for the heavy 
quarks, we use asymptotic or ``invariant'' quark masses
directly proportional to the Higgs v.\,e.\,v.\ times a Yukawa coupling.

In the limit of isospin symmetry we have 
\[ \Delta \ln \frac{m_p}{\hat{m}} = m_p^{-1} \sigma_{\pi N}(0) = 0.048\pm 0.01
\]
where $\hat{m}=(m_u+m_d)/2$, using $\sigma_{\pi N}(0)= 45\pm 10\,$MeV~\cite{GasserLS,BorasoyM}. Isospin-violating effects can be shown to be small 
by considering pion-nucleon scattering~\cite{MeissnerSteininger} or $m_p-m_n$,
hence we use the coefficient $0.048$ for both $u$ and $d$ quarks.

The strange contribution is related to the strangeness content of the nucleon
$y$ as   
\[ \frac{m_s}{m_p} \frac{\partial m_p}{\partial m_s}
= \frac{m_s}{\hat{m}}\frac{y}{2} \frac{\sigma_{\pi N}(0)}{m_p},\ \ 
y \equiv \frac{2 \vev{p|\bar{s}s|p}}{\vev{p|\bar{u}u+\bar{d}d|p}}. \]
An estimate $\partial \ln m_p/\partial \ln m_s\simeq 0.2$ was made 
in~\cite{FlambaumS} based on the lattice result $\vev{p|\bar{s}s|p}\simeq 
1.5$ with a sea-quark mass of $0.154\,$MeV~\cite{Dong}, corresponding to 
$y=0.36\pm 0.03$. However this large value was contested in~\cite{UKQCD}
where negative $y$ consistent with zero was obtained (negative values are 
unphysical). A safe but imprecise estimate is thus to take 
$\partial \ln m_p/\partial \ln m_s = 0.12 \pm 0.12,$ corresponding to 
$y= 0.2\pm 0.2$~\cite{BorasoyM}.

The electromagnetic contribution, composed of the direct electromagnetic 
self-energy and a part originating from nondegeneracy of 
pion masses~\cite{MullerMeissner}, gives
$m_p^{-1}\partial m_p/\partial\alpha \sim 1.8\times 10^{-3}$. 
Except in the case of nucleosynthesis, 
%where the neutron-proton mass difference is important, 
isospin-violating and electromagnetic effects can be neglected.\footnote{For 
a consistent unit-free treatment of $m_n-m_p$ in nucleosynthesis allowing 
both $\alpha$ and QCD parameters to vary, see~\cite{Langacker}; see also
\cite{Wetterich02} on implications for composition-dependent forces.} 
For consistency we must put 
\[ \frac{\Lambda}{m_p} \frac{\partial m_p}{\partial\Lambda} = 
1 - \sum_{q=u,d,s} \frac{\partial\ln m_p}{\partial\ln m_q} = 0.78 \pm 0.1. \] 
Putting it all together, we have 
\bea
\Delta \ln\frac{m_p}{M} &\approx& 0.78 \left( 
 \frac{2\pi}{9\alpha_3(M)} \frac{\Delta \alpha_3(M)}{\alpha_3(M)} +
 \frac{2}{27}\sum_{c,b,t} \Delta \ln\frac{m_q}{M} \
 + \frac{4}{9} \Delta\ln \frac{\tilde{m}}{M} \right) 
 \nonumber \\
 &+& (0.12\pm0.1) \Delta \ln\frac{m_s}{M} 
 + 0.048 \sum_{u,d} \Delta \ln\frac{m_q}{M} + \cdots \nonumber \\
 &=& \frac{0.54}{\alpha_3(M)}\Delta\ln \alpha_3(M) + 0.058 \sum_{c,b,t} 
 \Delta\ln \frac{m_q}{M} + 0.35 \Delta\ln \frac{\tilde{m}}{M} \nonumber \\
 &+& (0.12\pm0.1)  \Delta \ln\frac{m_s}{M} 
 + 0.048 \sum_{u,d} \Delta \ln\frac{m_q}{M} + \cdots \label{eq:14}
\eea
where apart from the strange term the individual coefficients have errors of
order $10\%$.
Now as before we can take the simplest unification scenario in which all 
superpartner and quark masses vary approximately with $v_H/M_X$ to obtain
\bea
\Delta \ln\frac{m_p}{M} &\approx& \left( \frac{0.54}{\alpha_X}  
 + (0.39\pm 0.12)\beta_v\ + 0.35\beta_S \right) \Delta \ln\alpha_X 
\label{eq:2.15} \\
 &\simeq& \left(13 + (13\pm 4) + (12\pm 1) \right) \Delta \ln\alpha_X \nonumber
\eea
where the last term on the R.H.S. is to be discarded in the absence of 
superpartners. Thus the inclusion of light quarks and SUSY thresholds and 
consequent reduction in the $\Lambda_c$ contribution has a non-negligible 
effect.
Using $\Delta \ln\mu/\Delta \ln\alpha_X \simeq 
(\Delta \ln (m_p/M_X)-\beta_v\Delta \ln\alpha_X)/\Delta \ln\alpha_X$ we 
come to the result
\beq
 \frac{\Delta \ln \mu}{\Delta \ln \alpha}\equiv \bar{R} \approx
 \frac{0.54 \alpha_X^{-1} + (-0.61\pm 0.12)\beta_v +0.35\beta_S }
 {0.022 \alpha_X^{-1} + 0.0018\beta_v + 0.011\beta_S } 
\eeq
Without variation in the weak and superpartner scales (setting 
$\beta_v=\beta_S=0$) we arrive at the rather model-independent expectation
$\bar{R}= 25\pm ({\rm few})$. This differs somewhat from the expectation 
$\bar{R}\simeq 36$ based on $m_p = {\rm const.}\times\Lambda_c$ and 
ignoring the effect on $\alpha$ of varying $\Lambda_c$ through light quark 
thresholds. (However we will see that this scenario is still ruled out.)

If we allow the weak scale to vary but ignore SUSY thresholds, setting
$\beta_v\simeq 34$, $\alpha_X=1/24$ and $\beta_S=0$, we obtain 
%$\bar{R}=-13.3 \pm 7$. 
$\bar{R} = -13 \pm 7$. The change here is mainly due to the variation in
$m_e/M_X$. If we now change $\beta_S$ to $34\pm2$ we arrive at 
%$\bar{R}=4.4 \pm 5$. 
$\bar{R} = 4 \pm 5$. Here, $\Delta\ln\mu$ turns positive and 
$\Delta\ln\alpha$ becomes somewhat larger through the variation of 
SUSY thresholds.

The ``varying $M_{G}$'' scenario of~\cite{Calmet2} for which $\alpha_G$ 
is constant can be recovered by taking the limit $\beta_v\rightarrow \infty$. 
In the presence of one should send $\beta_S$ to infinity and also enforce 
$\beta_v=\beta_S$. The expectation in this case is for $\bar{R}=-330\pm 65$ 
without superpartner contributions, or $\bar{R}=-20\pm 9$ with. Again, these 
differ from previous estimates~\cite{Calmet2} due to our more detailed 
treatment of $m_p$ and of light quark thresholds. The enormous value of 
$\bar{R}$ in the ``non-SUSY'' case is due to a cancellation 
%between charged matter and $W$ boson thresholds 
in $\Delta\alpha$ in the absence of the ``direct'' contribution.

The string-inspired scenario of~\cite{Dine} is more complex since it posits
a string scale $M_{\rm st}$ at which the gauge coupling is invariant, as 
well as an effective GUT scale $M_G$ varying relative to $M_{\rm st}$. 
SUSY-breaking and EWSB are taken to be independent of $M_G/M_{\rm st}$. 
There are two ways to implement the scenario: firstly, one may identify 
$M_X$ with $M_{\rm st}$ and set $\beta_v=\beta_S=0=\alpha_X$; however one 
has to rederive Eqs.~(\ref{eq:3}--\ref{eq:5}), (\ref{eq:12}--\ref{eq:13}), 
{\em etc.}, explicitly including heavy GUT thresholds whose contributions
$\propto b_i^{th}\Delta\ln(M_G/M_{\rm st})$ can be found given the 
beta-function of the unified group.

The second way is to identify $M_X\equiv M_G$ and then impose 
$\beta_v=\beta_S$
$= -\Delta \ln(M_G/M_{\rm st})/\Delta \ln \alpha_G$ (since the weak and 
supersymmetry scales are invariant relative to $M_{\rm st}$). The variation
in $\alpha_G$ is found by RG running down from $M_{\rm st}$ to $M_G$.
The results would differ slightly from those of~\cite{Dine}, again due to
different treatment of $m_p$ and light quark effects, but the overall 
conclusion that (even in the absence of rapidly-varying weak and SUSY scales)
the great majority of possible GUTs result in values $|\bar{R}|>10$, 
should remain.

\section{Varying radii of extra dimensions}
\subsection{Rationale of a varying radius}
In a model which is described by ($4+\delta$)-dimensional field theory 
over some range of energies, there is a hierarchy between the (inverse) 
compactification radius or radii $R_i^{-1}$ and the ultraviolet cutoff of 
the higher-dimensional theory $\Lambda_D$, where $D=4+\delta$. This makes 
it a well-defined problem to compute the effect of varying radius on the 
4D low-energy theory, since one can take physics above the scale 
$R_i^{-1}$ to be unchanged by such variation (apart from the change in 
masses of Kaluza-Klein modes), consistent with decoupling. It is 
conceivable that physics above this energy scale is also varying, but such 
variation would likely not be calculable since it might involve physics at 
energies equal to or above the cutoff. 

The superstring dilaton and moduli are {\em a priori}\/ candidates for 
cosmologically varying scalar fields and may be light enough to survive 
at low energies, yet they parameterize inverse radii which may be in the 
range $10^{16-18}\,$GeV. In principle they might come into our analysis, 
except that the theory above the inverse radius scale may be ``stringy'' 
to a significant extent: the cutoff is at the string scale (or near the 
11-dimensional Planck scale for M-theory), hence if the inverse radius 
also approaches this scale there is no range of energy over which 
extra-dimensional field theory makes sense. In perturbative string theory 
the dilaton- and moduli-dependence of low-energy quantities are 
extractable, but we face the usual problem that string models which are 
realistic enough to contain a reasonable approximation to the SM 
spectrum, are too complicated for detailed calculations to be practicable. 
One can however make simplifying assumptions based on the behaviour of 
less realistic models, leading to results such as those discussed in 
\cite{C+O,us,Dine}. 

\subsection{Are long-range forces inevitable? 
%\footnote{We thank a referee for highlighting this question.} 
}\label{sec:longr}
Na{\" \i}vely, one might expect that a varying radius which was the source
of varying alpha would mediate observable long-range forces. The reasoning 
is that, given that the scalar field appearing in the four-dimensional 
effective action (``radial modulus'') is evolving in a potential on 
timescales of order the inverse Hubble rate, its mass must be comparable to 
the Hubble rate, thus it mediates a force over very large distances. The 
strength of the force ({\em i.\,e.} its coupling to matter) in Kaluza-Klein 
type theories is found at tree level to be comparable to gravity (see 
{\em e.\,g.} \cite{DimopoulosG}) thus it is likely ruled out by Solar 
System tests of deviations from general relativity and tests of the Weak 
Equivalence Principle \cite{Will01}.~\footnote{Even if one is allowed to 
adjust by hand the scalar kinetic term and couplings to matter, if one 
specifies that it should be the origin of the alpha variation, then 
apparent violations of the WEP should be not far below current bounds 
\cite{DvaliZ,Wetterich02}.}
(Of course one would not observe any deviation from the inverse square
law in the laboratory.)

However, this reasoning is based on a number of assumptions which are not
universally valid: therefore although it appears difficult for a varying 
radius to evade these bounds, it cannot be dismissed immediately. 

The first assumption is that the apparent measured variation is due to a
smooth evolution over billions of years. However, as we mentioned 
in Section \ref{sec:other} the variation may be spatial and long-range or
short-range; the variation may even operate on distance scales as small 
as a galactic halo. Hence the mass of a varying scalar may be some orders 
of magnitude above the inverse Hubble radius. This does not much help to 
evade observational bounds, but it is worthwhile to refute an unwarranted 
assumption.

The second is that the radial modulus is identified with the light scalar 
that is slowly rolling in a non-varying potential. However, compactified 
radii may vary without the modulus being light. There is an analogous 
argument for two scalar degrees of freedom in the SM, the sigma meson 
corresponding to excitations of the chiral condensate and the Higgs. In the 
previous parts of this paper we saw that these scalar v.\,e.\,v.'s are 
likely varying (in Planck units) in a way correlated with alpha. By the 
same argument one would also conclude that the sigma and the Higgs should 
have masses of order the inverse Hubble length. In fact, the scalars are 
massive and always at the minima of their potentials: their variation is 
induced by a varying potential. There is no reason why a variation of the 
radius should not similarly be induced by a varying stabilization potential, 
with the mass of the modulus being $10^{-3}\,$eV (the current limit of 
sensitivity of fifth-force experiments) or greater. We do not specify the 
source of such a varying potential, except that it should not directly 
affect the coupling constants of SM fields. Thus the radius may 
{\em mediate}\/ the variation to the observable sector.

The third is that the scalar couplings to matter are constant. However, 
even in the simplest case of unification, this is not the case, since any 
variation in the SU$(3)$ coupling at high scale results in nonperturbative 
dynamics at low energy which determines the masses of nucleons as an 
exponential function. When one combines this fact with the presence of 
Scherk-Schwarz SUSY-breaking, electroweak symmetry-breaking, varying Yukawa 
couplings and other low-energy phenomena it becomes clear that the effective 
couplings to matter will be field-dependent, possibly in a complicated way. 
It was shown in \cite{DamourP} that, in the case of field-dependent couplings 
to matter, if there is a value of the scalar where the coupling is small, 
this value is an attractor for the evolution of the scalar. Hence the 
deviations from GR may be suppressed. This mechanism was discussed in a 
model with varying radius in \cite{Albrechtquint}, although the details of 
this model are different from those I discuss below.

The fourth is that the tree-level action for the radial modulus is a good 
approximation. However, it is possible that there are large corrections
to the kinetic term and couplings. In \cite{Albrechtquint} radiative 
corrections to the action were mentioned as a possible way to evade 
bounds on a light scalar.

Violations of the WEP and other deviations from GR are very strong bounds 
on {\em all}\/ models which seek to explain the dynamics of varying alpha, 
not only models with varying radii. We do not have any definite solution 
to the problem: clearly the question is to be addressed by looking in 
detail at different models. In considering correlations between observables 
at a given epoch or in similar environments, the details of the radial 
modulus or other light scalar dynamics are not directly relevant.

\subsection{Scherk-Schwarz revisited}
As mentioned in the Introduction, Scherk-Schwarz symmetry breaking, in
combination with varying radii, appears to lead to a relation between 
perturbative gauge couplings and the mass scale of symmetry-breaking very
different from radiative or nonperturbative sources of SSB.

However, there are subtleties in the analysis which need to be addressed. 
It has recently been shown that Scherk-Schwarz breaking in a wide variety 
of situations can be reformulated as spontaneous breaking with v.\,e.\,v.'s 
being nontrivial functions of position in the extra dimension(s) 
(see {\em e.\,g.}~\cite{Bagger:2001qi}). This may blur the distinction 
between Scherk-Schwarz breaking and breaking by radiative or 
nonperturbative means. Various connections have been made between 
Scherk-Schwarz supersymmetry-breaking in the fifth dimension and nonzero 
v.\,e.\,v.'s of auxiliary fields which might arise dynamically, both in
heterotic M-theory~\cite{AntoniadisQ} and in simpler orbifold contexts 
\cite{Marti:2001iw}.
%In one case, breaking $\mathcal{N}=2$ supersymmetry in 5 dimensions by 
%imposing by hand boundary conditions on an $S^1/Z_2$ orbifold, is 
%equivalent to an $x^5$-dependent v.\,e.\,v.\ for an auxiliary field in $D=5$ 
%supergravity (see {\em e.\,g.}~\cite{QuirosG}). 

Such a connection might justify the choice of an extremely small 
dimensionless parameter $\tilde{\alpha}\sim 10^{-13}$ in~\cite{Nomuradesert} 
to characterize the ``twisting'' of the boundary conditions. Thus what 
originally appeared as a geometric and explicit breaking induced by a
nondynamical choice of boundary conditions, turns out to be a dynamical 
spontaneous breaking whose size may be determined by nonperturbative 
gauge theory effects. (In both cases the breaking is soft.)

Hence, it is crucial to trace the mechanisms of symmetry-breaking back to 
their source, since consistent predictions for varying couplings depend on 
knowing how the mass scales and v.\,e.\,v.'s of the model are generated. In 
the case just mentioned, since gaugino condensation would take place at a 
scale much lower than that of the inverse radius, we must take account of the 
variation in the condensate resulting from a variation of the radius. Thus 
the assumption that the S-S parameter $\tilde{\alpha}$ is constant, which 
would imply that the SUSY-breaking masses vary strictly with $1/R$, is 
likely incorrect. However, if a S-S breaking parameter is a number of order 
1 or quantized, we take it to be constant since it does not require a 
separate dynamical explanation. (See~\cite{vonGersdorff:2002tj} for a 
discussion of dynamics that could fix the value of such a parameter).

\subsection{Units and definitions}
For varying extra-dimensional radii, the relevant dimensionless 
quantities are the radius relative to the cutoff of the extra-dimensional 
field theory $R\Lambda_D$, and relative to the extra-dimensional Planck 
length $RM_P^{(D)}$. Since by assumption $(4+\delta)$-dimensional physics 
above $R^{-1}$ is invariant, we expect $\Lambda_D/M_P^{(D)}$ to be 
unchanged, hence the two ratios carry the same information. We will not be 
dealing with gravitational effects, hence we take $R\Lambda_D$ as the
``order parameter''.

In contrast to the usual designation of $M_P$ as a constant, resulting in
``Planck units'' if we have $M_P=1$ (other conventions import factors of 
$\sqrt{8\pi}$, {\it etc}.), we could choose to work in ``radius units''
in which $R$ is constant and $\Lambda_D$ and $M_P$ vary. Thus 
``varying $R$'', in addition to affecting the classical relation between
couplings in $4+\delta$ dimensions and $D=4$, implies RG evolution in the 
extra-dimensional theory, since we are changing the ratio of the UV scale 
to the IR scale(s). Since there are a large number of thresholds in such 
theories, with masses determined primarily by $R$, radius units may be 
more convenient. 
%Apart from the effect on gravitation, the result of a varying cutoff is that couplings evolve over a different energy range and take different values on hitting the scale $R^{-1}$ (or any of the associated Kaluza-Klein modes). 

\subsection{RG ``running'' in $4+\delta$ dimensions}
As is well known~\cite{DDG}, the behaviour of coupling constants under 
change of scale is radically altered above the compactification scale if 
fields propagate in extra dimensions. The usual procedure is to decompose 
extra-dimensional operators into K-K modes whose masses are fixed multiples 
of $R^{-1}$ and treat the massive modes as thresholds at which the effective 
four-dimensional beta function changes. With a large number of such modes, 
one finds an ``averaged-out'' behaviour of power-law running of the effective 
4d coupling. Obviously, the running depends on which fields are allowed 
to propagate ``in the bulk'', and proceeds as normal in 4d if all fields
coupled by a particular operator are ``on the brane''. 

A cutoff is required since gauge and Yukawa couplings quickly become 
nonperturbative with increasing energy, being irrelevant operators in 
higher dimensions~\footnote{For energies far enough above $R_i^{-1}$, 
the behaviour of the compactified theory should reproduce that in 
uncompactified $D=4+\delta$ field theory, and the Kaluza-Klein modes
can be approximated by a continuum.}.
For boundary conditions imposed at a scale $\Lambda_D$ to be meaningful, 
the theory should be perturbative at this scale. By assumption, the 
coupling strengths of all operators in the $(4+\delta)$-dimensional 
theory at and above $\Lambda_D$ do not vary.

We quote the relevant formula from~\cite{DDG}
\beq \label{eq:17}
\alpha_i^{-1} (\mu^-) = \alpha_i^{-1} (\Lambda_D) - 
\frac{b_i}{2\pi}\ln \frac{\mu^-}{\Lambda_D} 
+ \frac{\tilde{b}_i}{2\pi}\ln\frac{M_\delta}{\Lambda_D} 
+ \frac{\tilde{b}_i X_\delta}{2\pi \delta_i}\left(\left(\frac{\Lambda_D}
{M_\delta}\right)^{\delta_i} -1\right)
\eeq
where $M_\delta=R^{-1}$ is the mass of the first Kaluza-Klein mode. Thus
\beq \label{eq:18}
 \frac{\Delta\ln \alpha_i(\mu^-)}{\alpha_i(\mu^-)} = 
 \left(\delta_i \alpha_i^{-1}(\Lambda_D) -\frac{\tilde{b}_i}{2\pi}
-\frac{X\tilde{b}_i}{2\pi} 
 \left(\frac{M_\delta}{\Lambda_D}\right)^{-\delta_i}
\right) \Delta \ln \frac{M_\delta}{\Lambda_D} + \frac{b_i}{2\pi}\Delta \ln 
\frac{\mu^-}{\Lambda_D},
\eeq
recalling that $\alpha_i (\Lambda_D) \propto 
(M_\delta/\Lambda_D)^{\delta_i}$,
where the $i$'th gauge group propagates over $\delta_i$ extra
dimensions. The first term inside brackets arises from this classical
dependence, the second from logarithmic K-K contributions and the third
from the power-law correction; the last term being the usual $D=4$ 
scale dependence. Formally, the second and third terms inside brackets are 
suppressed by a loop factor relative to the first, however the power-law 
term may have a large enhancement, essentially due to a large number of 
K-K modes. However, we can choose the cutoff such that the classical term 
dominates, as follows.

If couplings are perturbative at the cutoff $\Lambda_D$, we can choose a 
lower effective cutoff $\Lambda'$, integrating out all 
modes between the two. For $\Lambda'>M_{\delta}$ the form of Eq.~(\ref{eq:17}) 
will remain the same and the extra-dimensional gauge coupling at 
$\Lambda'$ can equally be taken constant. Thus we lower $\Lambda'$ to just 
above $M_{\delta}$, making the power-law contribution of K-K modes as small as 
possible and, in practice, negligible. The quantum terms in the 
brackets can then be neglected as formally of higher order, since we do not
expect $\tilde{b}$ to be very large. We similarly neglect (the variation in)
the difference between $\ln (\mu^-/\Lambda')$ and $\ln (\mu/M_\delta)$, and
set $\alpha_i(\Lambda')\simeq\alpha_i(M_\delta)$ in all {\em static}\/ 
quantities. Whether $(b_i/2\pi)\Delta \ln (\mu^-/M_\delta)$ is also of 
higher order, will depend on how the low energy scale is dynamically 
generated: if $\mu^-$ varied with the QCD scale $\Lambda_c$ it could compete 
with the first term.

Similar considerations for Yukawa couplings imply that the simple
classical formulae
\[
\Delta \ln \alpha_i(M_\delta) = \delta_i \Delta\ln 
 \frac{M_\delta}{\Lambda'},\ \ \
\Delta \ln y(M_\delta) = \left(\frac{p_y}{2}+c\right) 
 \Delta \ln \frac{M_\delta}{\Lambda'}
\]
are good approximations in the perturbative regime, where the Yukawa 
coupling $y$ couples $p_y=0,1,2,3$ fields (either fermionic or scalar) 
propagating in extra dimensions and $(3-p_y)$ fields localized 
``on the brane''. The integer $c$ takes the value $0$ for Yukawa couplings 
localized in extra dimensions, as in some models based on $N=2$ 
supersymmetry in $D=5$, and $1$ for a bulk coupling in the case where all
fields propagate in the extra dimension. For a supersymmetric Higgs mass
term $\mu_SH_1H_2$ localized on the brane with Higgs in the bulk, $\mu_S$ 
varies directly proportional to $M_\delta$, $\mu_S/M_\delta=$constant; for 
other configurations ({\em i.\,e.} universal extra dimensions or localized
Higgses) $\mu_S$ arises directly from a mass term in the $D=5$ theory and 
thus varies as $\mu_S/M_\delta \propto (M_\delta/\Lambda')^{-1}$.
%~\footnote{In radius units, any coupling derived from a mass term in extra dimensions varies relative to the compactification scale $M_\delta$.}

We take one varying dimension, the generalization to more than one is 
trivial. Substituting for $\alpha_i(M_\delta)$ and neglecting the 
variation in low-energy thresholds, including $\mu^-$, relative to 
$M_\delta$, we find
\beq \label{eq:19}
\Delta \ln \alpha_i(\mu^-) = \frac{\alpha_i(\mu^-)}{\alpha_i(M_\delta)}
\delta_i \Delta \ln \alpha_i(M_\delta) = 
\left(1+\frac{\alpha_i(\mu^-)b_i}{2\pi} \ln \frac{\mu^-}{M_\delta}\right) 
\delta_i \Delta \ln \frac{M_{\delta}}{\Lambda'}.
\eeq
Then using the na{\" \i}ve formulae $\Delta \alpha^{-1} =
\Delta \alpha_Y^{-1} +\Delta \alpha_2^{-1} $ and 
$\Delta \ln (m_p/m_e) = (2\pi/9\alpha_3(M_\delta))$ $\Delta \ln \alpha_3(M_\delta)$, and neglecting running of $\alpha$ below the electroweak scale and 
possible variation in $m_e/M_{\delta}$, we find
\bea 
\Delta \ln \alpha &=& \left(\delta_Y\cos^2 \theta_W + \delta_2\sin^2 \theta_W
 + \frac{\alpha(M_Z)}{2\pi}(\delta_Y b_Y+\delta_2 b_2)
 \ln \frac{M_Z}{M_\delta} \right) \Delta \ln \frac{M_{\delta}}{\Lambda'}, 
 \nonumber \\
\Delta \ln \mu &=& \frac{2\pi\delta_3}{9}\left(\alpha^{-1}_3(M_Z) + 
 \frac{b_3}{2\pi}\ln \frac{M_Z}{M_\delta}\right) \Delta\ln
 \frac{M_{\delta}}{\Lambda'} 
\eea
thus without superpartners we have
\beq \label{eq:21}
\bar{R} \approx \delta_3\left(5.9 -\frac{7}{9}\ln \frac{M_Z}{M_\delta} 
 \right)
 \left(\delta_Y\cos^2 \theta_W + \delta_2\sin^2 \theta_W 
 + \frac{1}{804}\left(\frac{41}{6} \delta_Y - 
 \frac{19}{6} \delta_2\right)\ln \frac{M_Z}{M_\delta} \right)^{-1}
\eeq
and with superpartners (in the case $M_\delta \gg \tilde{m}$)
\beq \label{eq:22}
\bar{R} \approx \delta_3\left(5.9 -\frac{3}{9}\ln \frac{M_Z}{M_\delta} 
 \right)
 \left(\delta_Y\cos^2 \theta_W + \delta_2\sin^2 \theta_W 
 + \frac{1}{804}\left(11 \delta_Y + 
 \delta_2\right)\ln \frac{M_Z}{M_\delta} \right)^{-1}.
\eeq
Enforcing $\delta_Y=\delta_2=\delta_3$ as required by an 
extra-dimensional GUT we obtain 
\beq \label{eq:23}
\bar{R} \approx \frac{ 5.9- \frac{7}{9}\ln (M_Z/M_\delta) }
 { 1+\frac{1}{804}\cdot\frac{11}{3} \ln (M_Z/M_\delta) }\ \mbox{[non-SUSY]},\
\frac{ 5.9- \frac{3}{9}\ln (M_Z/M_\delta) }
 { 1+\frac{1}{804}\cdot 12 \ln (M_Z/M_\delta) }\ \mbox{[SUSY]}.
\eeq
If we use the more detailed treatment of the proton mass Eq.~(\ref{eq:14})
then the value of $\bar{R}$ is simply reduced by a factor $0.78$. 

In the small radius limit $M_\delta\rightarrow M_G$ where the K-K modes 
do not influence unification we recover the SUSY-GUT expectation 
$\bar{R}\approx 33$. In the opposite limit where $M_\delta$ approaches 
$M_Z$ we would obtain $\bar{R} \gtrsim 6$. 
%(or $4.5$ with the more accurate proton mass formula).
% $5.9$. 

The suppression of $\bar{R}$ compared to its value in 4D GUTs was named 
$\kappa$ in~\cite{PC_S_T}, who found $\kappa\simeq 0.16$ in the limit of 
``low scale unification''. Our estimate is $\kappa\gtrsim 0.18$ taking the 
value $\bar{R}_{\rm 4D GUT}\approx 33$. 
%or $\kappa\gtrsim 0.16$ 
Note that the estimate $\bar{R}_{\rm 4D GUT}\sim 36$ used in~\cite{PC_S_T}
can also be obtained from our formula, after accounting for the running of 
$\alpha_{\rm em}$ below $M_Z$ through $\Delta \ln \alpha = 
(\alpha/\alpha(M_Z))\Delta\ln \alpha(M_Z)$.
% $35.5+$

The above expressions for $\bar{R}$ are obtained without knowing {\em any 
details}\/\ of the supposed extra-dimensional GUT theory: the identity of 
the GUT group, the pattern of breaking, the localization of matter, 
{\em etc.} We do not even need to have a unified theory! The only parameter 
is the ratio $M_Z/M_\delta$.~\footnote{Of course, if unknown intermediate thresholds appear below $M_\delta$, all predictions are off.} 
Taking a value $M_\delta=5\,$TeV slightly larger than the minimum, without 
superpartners, we obtain $\bar{R}\approx 9.2$:
%(or $7.6$ with the more accurate proton mass formula): 
the logarithmic variation of $\alpha_3^{-1}(M_\delta)$ with increasing 
$\alpha_3^{-1}(M_\delta)$ is significant at low compactification scales.

In more general ``brane world'' models, one has more freedom to choose
the integers $\delta_i$ and $p_y$. Clearly, the choice $\delta_3=0$ leads
to $\Delta\ln \mu=0$, up to effects (so far set to zero) of thresholds 
below $M_\delta$ and variation in $m_e/M_\delta$. This is just the obvious 
scenario in which the strong force does not ``feel'' the variation because 
it is not propagating round the varying dimension. Data will turn out to 
favour small values of $\bar{R}$, thus assigning $\delta_3=0$ is an obvious 
choice, if somewhat arbitrary in the absence of a concrete model. 

To assess the influence of thresholds and varying $m_e/M_\delta$ we will 
now turn to more concrete models of mass generation (which, however, will
not have the feature that $\delta_3=0$).

\subsection{Electroweak breaking by boundary conditions}
The possibility of electroweak symmetry-breaking triggered by 
Scherk-Schwarz supersymmetry-breaking on an orbifold~\cite{Quiros,Nomurawk} 
leads to a class of predictive scenarios where the variation of all 
observables, under a variation in radius, can be found with very 
few assumptions. The variations in gauge and Yukawa couplings (and $\mu_S$) 
as sketched above can be applied to obtain the behaviour of $v_H$ and 
fermion masses. The main result will be the variation of $m_e/M_\delta$
and the resulting effect on $\mu$. As we will see, given the assumption that
all mass parameters in the electroweak sector originate from compacification, 
the fractional variation in fermion and vector boson masses relative to 
$M_\delta$ cannot be more than a few times that in $M_{\delta}/\Lambda'$ 
(compared to around thirty times, in the scenarios of high-scale models 
with EWSB through dimensional transmutation). Thus variations in 
charged thresholds are indeed small enough to neglect in these models. As 
for supersymmetric partners, the nature of SUSY-breaking ensures that the 
leading behaviour of soft masses is a constant times $M_\delta$, thus the 
variation of their thresholds is also negligible. Since (depending on details 
of the model) superpartner masses are around the compactification scale, 
we use the SM beta-functions to run up to $M_\delta$ 
(as in Eq.~(\ref{eq:19})). 

As a first estimate, we truncate the Higgs potential to the quadratic and
quartic terms and assume that first, the quadratic Higgs mass term is 
generated solely by the radiative effects of the top and stop loops
and thus varies as $M_\delta^2 y_t^2$, and second, that due to the 
underlying SUSY structure the quartic Higgs couplings vary as 
$\alpha_2(M_\delta)+\alpha_1(M_\delta)$. 

Thus, in the one-Higgs model of~\cite{Nomurawk}, in this approximation, the  
Higgs v.\,e.\,v., which varies as $(|m_H^2|/\lambda)^{1/2}$, behaves as 
$\Delta \ln (v_H/M_\delta) \approx \Delta \ln (M_\delta/\Lambda')$ and 
fermion masses vary as $\Delta \ln (m_{q,l}/M_\delta) \approx (5/2) \Delta 
\ln (M_\delta/\Lambda')$.
In the two-Higgs model of~\cite{Quiros} the Higgs v.\,e.\,v.\ is directly 
proportional to $M_\delta$, $\Delta \ln (v_H/M_\delta) \approx 0$, and 
the quark and lepton masses have a weak dependence $\Delta \ln 
(m_{q,l}/M_\delta) \approx (1/2) \Delta \ln (M_\delta/\Lambda')$. The
difference is simply due to different localization of Standard Model fields.

\subsubsection{Detailed treatment of the Higgs potential}
One may treat the models more precisely by using the full Higgs potential,
which may include terms besides the usual quadratic and quartic ones. 
In~\cite{Nomurawk} the resulting dependence is $v^4 = ({\rm const.})
(g^2+g'^2)^{-1}R^{-4}$: the expected dependence on $y_t$ does not 
appear, due to a quirk of minimization of the potential, and the 
dependence on the quartic D-term coupling is different. Hence we have 
$\Delta \ln (v_H/M_\delta) \simeq (-1/2) \Delta \ln (M_\delta/\Lambda')$ 
and $\Delta \ln (m_{q,l}/M_\delta) \simeq \Delta \ln (M_\delta/\Lambda')$. 

In the model of~\cite{Quiros} the minimization conditions do not give 
a simple form for the dependence of $v_H$ on input parameters, due to the
presence of two Higgs doublets and radiative corrections. The ratio of 
the two Higgs v.\,e.\,v.'s $\tan\beta$ is around $37$ so we neglect 
$1/\tan^2\beta$ next to 1 and obtain
\beq \label{eq:24}
\Delta \ln \frac{M_\delta}{\Lambda'} = \Delta \ln \left\{
  \left(\frac{\mu_S}{M_\delta}\right)^2 + 
  \left( 2\lambda_t + \frac{g^2}{8\cos^2 \theta_W} \right)
  \left(\frac{v_H}{M_\delta}\right)^2 \right\} 
%+ {\mathcal O}(1/\tan^2\beta)
\eeq
where $\lambda_t$ is a quartic coupling proportional to $y_t^4$ with a 
weak logarithmic dependence on $RM_Z \propto M_Z/M_\delta$. Then clearly 
the dependence $\mu_S/M_\delta\propto (M_\delta/\Lambda')^{-1}$ is the 
dominant effect for the range of values of $\mu_S$ considered in 
\cite{Quiros}. For Scherk-Schwarz
parameter $\omega=1/3$, $\mu_S=600\,$GeV and $R^{-1}=9\,$TeV we find 
$\Delta\ln (v_H/M_\delta) = 44 \Delta\ln (M_\delta/\Lambda')$ while for 
$\mu_S=200\,$GeV and $R^{-1}=3\,$TeV we find $\Delta\ln (v_H/M_\delta) = 
7 \Delta\ln (M_\delta/\Lambda')$. Roughly, $\Delta\ln (v_H/M_\delta)$ 
scales with $(\mu_S/M_Z)^2$.

The extreme sensitivity to $\mu_S$ and the fact that this parameter is put in
by hand prevent the model from making any definite prediction. Essentially
there is no dynamical explanation of the mu-term, it is a brane mass term 
in the $D=5$ theory which we have to assume is non-varying relative to the 
extra-dimensional cutoff. It is not correlated with the compactification 
scale (times some coupling constants), thus, in radius units, it varies 
inversely compared to the radiative corrections triggering EWSB. If 
$\mu_S$ is somewhat larger than $M_Z$ it completely dominates the 
variation of $v_H$ giving a large and arbitrary result. 

In general, if there are comparable mass scales with different dynamical 
origins, the variation of a quantity obtained by adding different mass 
terms can be very sensitive to the proportion of different terms, thus 
the model parameters need to be known rather precisely. If the mu-term 
were to arise from the compactification scale then the dependence on the 
actual value of $\mu_S/M_Z$ would be much milder and the scenario could 
be predictive.

\subsubsection{More precise estimate of $\bar{R}$}
Finally we combine the variation of $v_H$ and fermion masses in the model
of~\cite{Nomurawk} with the detailed treatment of the proton mass to obtain
the estimate of $\bar{R}$. In this case the variations in $m_f/M_\delta$ is
indeed small enough that one can neglect all threshold corrections and
quark mass contributions to the variation of $m_p/M_\delta$. Taking
account of the variation  
$\Delta \ln (m_e/M_\delta) \simeq \Delta \ln (M_\delta/\Lambda')$,
which simply subtracts 1 from the denominator of Eq.~\ref{eq:23}, we find 
\[ \bar{R} = 6.3 \pm 1. \] 
It has been noted that a divergent Fayet-Iliopoulos term might arise in 
models of this type~\cite{F-I_term}: the question of whether this could 
have a considerable effect on $\bar{R}$ is interesting but beyond the
scope of this paper.

\subsection{GUT breaking by boundary conditions}
At the other end of the scale of compactification radii are scenarios where
the energy scale of the extra dimension is just below the SUSY-GUT scale, 
as required if a GUT is to be broken by orbifold boundary conditions 
\cite{Nomuragut,Kawamura}. These types of theory do not have dramatic 
low-energy predictions of detectable K-K modes, however they are claimed to 
be an improvement over four-dimensional GUTs in terms of predictivity and 
agreement with current values of $\sin^2 \theta_W$ and $\alpha_3(M_Z)$, at 
least for certain simple models. The ratio of the compactification scale to 
a fundamental cutoff may also play a r{\^ o}le in generating fermion mass 
hierarchies, thus the variation of fermion masses and $y_t$ with a varying 
radius may be under control (although a full model explaining the ratio 
$m_e/m_\tau$ has not appeared).

However, as previously noted the mechanism of EWSB is not fully accounted 
for in such models: the dynamical origin of the hierarchy $M_W/M_\delta$ is 
not specified. Since this is also the major cause of uncertainty in 
four-dimensional GUT models, extra-dimensional GUTs of this type currently
do not have predictions distinguishable from 4D GUTs for varying couplings.

\section{Comparison with data and conclusion} \label{sec:data}
There are four types of observable that may be considered in the redshift
range of interest: $\alpha$, $X \rightarrow \alpha^2g_pm_e/m_p$, 
$Y \rightarrow \alpha^2g_p$ and $\mu \rightarrow m_p/m_e$ where 
$\rightarrow$ signifies ``is a measurement of''. Current values are as 
follows:
\begin{eqnarray}
\alpha{-1} \Delta \alpha &=& (-0.543 \pm 0.116)\times 10^{-5},\ 0.2< z < 3.7\ \cite{Webb03} \\
X^{-1} \Delta X &=& (0.7\pm 1.1)\times 10^{-5},\ z=1.78\ \cite{CowieS} \\
Y^{-1} \Delta Y &=& (-0.16\pm 0.54)\times 10^{-5},\ z=0.68, \\
\ & \ & (-0.20\pm 0.44)\times 10^{-5},\ z=0.25\ \cite{Murphy01} \\
\mu^{-1} \Delta \mu &=& (5.0\pm 1.8)\times 10^{-5},\ z=3.02\ \cite{Ivanchik}
\end{eqnarray}
Theory and data can be usefully compared for $\alpha$ and $\mu$;
a variation in $g_p$ cannot reliably be related to more fundamental 
parameters~\footnote{It has been claimed that $g_p$ cannot vary 
substantially, being essentially a Clebsch-Gordan coefficient~\cite{Dine},
but also that it may have a substantial but unknown correlation with 
the strange quark mass~\cite{FlambaumS}.}. 

With three independent types of measurement and four observables, one may 
in principle check consistency and obtain values of $g_p$ or $m_p/m_e$ in 
two different ways. Given measurements at similar redshifts and ignoring 
complications due to spatial variation, one could eliminate $g_p$ as follows: 
\beq 
\frac{Y}{\alpha^2} \rightarrow g_p,\qquad 
\frac{X\mu}{\alpha^2} \rightarrow g_p.
\eeq
If the resulting fractional variation were consistent but large compared 
to $\alpha^{-1} \Delta \alpha$, it would indicate a large quark mass 
contribution to $g_p$, which would be rather unexpected. Similarly one 
can compare the ``direct'' measurement of $\mu$ to that extracted from 
\[
\frac{Y}{X} \rightarrow \frac{m_p}{m_e}
\]
which will be consistent if and only if the values for $g_p$ are. The 
value of $\alpha$ itself derives only from the ``direct'' measurement.

If we pretend that current data can legitimately be combined, then 
$Y/\alpha^2$ would give $\Delta \ln g_p = (0.93 \pm 0.6)\times 10^{-5}$ 
whereas $X\mu/\alpha^2$ gives $\Delta \ln g_p = (6.8\pm 2.1)\times 
10^{-5}$. The two values differ by over 2.5 standard deviations, so 
one hesitates to combine them; the second value also deviates from zero 
by over 3 sigma. 
The variation in $m_p/m_e$ derived from $Y/X$ would be $\Delta \ln \mu
 = (-0.9\pm 1.2)\times 10^{-5}$, also differing from the ``direct'' 
measurement by over $2.5\sigma$. But such combinations of measurements 
are strictly not correct, since they correspond to widely differing 
redshifts and environments. Such checks can only be taken seriously if 
the number and reliability of data increase significantly and the data 
sets truly overlap. 

However, both the measurement of $\mu$ at $z=3.02$ and the derived
value $Y/X$ at redshift around 1 bound any model where smooth and not too
rapid time variations in $\alpha$ imply variations in $m_p/m_e$, since
a nonzero variation in $\alpha$ appears to occur at least for redshifts 
1 to 3. To stand a chance of agreeing with experiment, $\Delta \ln \mu$ 
should be in the range $(-2.9,5.8)\times 10^{-5}$, thus $\bar{R}$ should 
be in the range $(-10.5,5.5)$ (rounded to the nearest half integer).

It is difficult for many of the models to fit in this window: the ``best 
fit'' among 4d unified theories discussed above is the case where 
$\beta_v$ and $\beta_S$ are around 34 (some scenarios discussed in 
\cite{Dine} may also survive). With some ``adjustment'' of $\beta_v$ to 
take account of a varying top Yukawa in concrete models of radiative EWSB
a better fit might be obtained, but this hardly constitutes a motivation 
for further investigation. Our main result is that in models with high-scale
unification the variation of thresholds and fermion masses can be a {\em 
leading}\/\ contribution to the variations in $\alpha$ and $\mu$: we 
presented a detailed discussion of these contributions.

In extra-dimensional models, the variation of a compactified dimension 
relative to the cutoff can lead to definite predictions, both for coupling
constants and mass ratios, even if there are no unification relations. One 
only needs to know the physics below the compactification scale. Predictions 
depend mostly on the compactification scale $M_\delta$ and on which fields 
propagate in extra dimensions. As a special case, we confirm a previous 
result~\cite{PC_S_T} that the variation of the strong coupling is suppressed 
relative to the variation of $\alpha$ in extra-dimensional GUTs with 
low unification scale. 

If, as in some ``brane-world'' models, the gluons of QCD do not propagate 
round the varying dimension, while one or both of the SU$(2)\times$U$(1)$ 
groups do, the variation of the RG invariant scale $\Lambda_c$ and the 
proton mass (relative to $M_\delta$) derive only from thresholds and quark 
masses and are likely to be small, while the variation in $\alpha$ is not 
suppressed. Thus one contribution to $\bar{R}$ is small. 
However, such a scenario also requires a concrete mechanism of EWSB to be 
self-consistent. If the variations of the electron and quark masses also 
turn out to be small, one would reach the region of $\bar{R}$ allowed by 
observation. 

Note that the coupling of a light scalar $\phi$, supposed to be the source of
the variation, to matter depends on the function $dm/d\phi$, where $m$
is the mass-energy of the body considered in Planck units. This quantity, 
which is correlated with some of the deviations from GR, is at leading 
order given by $d(m_N/M_P)/d\phi$. Violations of the WEP due to 
composition-dependent forces will also include a term varying as 
$(d/d\phi)(m_p+m_e-m_n)/M_P$. Thus, scenarios in which the variations of 
$m_p/M_P$ and $v_H/M_P$ are both small are better placed to satisfy bounds 
on both the variation of $\mu$ and on deviations from GR, than if 
$\bar{R}$ is small but $m_e/M_P$ and $m_p/M_P$ both have large variations.

In the extra-dimensional models we considered which do include a definite 
recipe for EWSB, all gauge bosons are in fact required to propagate round 
the compact dimension.
In one model, the variation of $v_H$ relative to $M_\delta$ was extremely
sensitive to the undetermined value of the mu-term $\mu_S$, thus we were not
able to arrive at a prediction. In another model with fewer parameters,
agreement with data is tenuous in that the prediction for $\bar{R}$ 
%obtained by lowering the compactification scale to near $M_Z$ 
is positive and larger than 5, even including the variation in 
$m_e/M_\delta$. However it is possible that slightly different models of EWSB 
might result in definite predictions with a larger variation in 
$m_e/M_\delta$ which could bring $\bar{R}$ within an acceptable range.

%\begin{table}%[H] add [H] to break across page

%\caption{Dependence of CKM parameters and $m_u$ on $\zeta$ for the
%two-parameter mass matrices, for two sets of random coefficients 
%$d^{u,d}_{ij}$.}
%\begin{ruledtabular}
%\centering
%\begin{tabular}{|c|ccccc|} \hline
%$\zeta$& $\delta_{\rm KM}$& $F_{CP}\equiv\delta_{\rm KM}/\zeta$& $J\times 10^5$& 
%$m_u(\mbox{MeV})$& $|V_{13}|\times 10^3$ \\ \hline
%$0.0001$ 

%& $-0.065$, $0.033$ %delta $0.0012$, $0.034$
%& $-650$, $340$ %d/z $12$, 
%& $0.16$, $-0.079$ %J.10^5 $-0.0029$, $-0.082$
%& $4.0$, $4.1$ %mu $5.5$, $3.7$
%& $2.2$, $2.2$ %v13.1000 $2.2$, $2.2$ 
%\\

%$0.0003$ 

%& $-0.19$, $0.100$ %$0.0036$, $0.103$
%& $-645$, $330$ %$12$, $340$
%& $0.47$, $-0.24$ %$-0.0086$, $-0.25$
%& $6.0$, $7.0$ %$12.8$, $3.8$
%& $2.3$, $2.2$ %$2.2$, $2.2$
%\\

%\hline
%\end{tabular}
%\end{ruledtabular}
%\end{table}

%\begin{figure}
%\centering
%\includegraphics[width=11cm,height=8.5cm]{Jcorr.eps}
%\caption{Amended version of Fig.~2 of~\cite{Lebedev}: the physical $J$
%of matter {\em vs.}\/\ antimatter.}
%\label{fig:corr}
%\end{figure}

\subsection*{Acknowledgments}
The author would like to thank Christof Wetterich, Keith Dienes, 
Stefan Scherer, Malcolm Fairbairn, Antonio Delgado and especially Yasunori 
Nomura for helpful correspondence.
Work supported by the EU Fifth Framework Network `Across the Energy 
Frontier' (HPRN-CT-2000-00148).
%\end{acknowledgments}


\end{document}